\documentclass[10pt, conference, compsocconf]{IEEEtran}

\ifCLASSINFOpdf

\else
 
\fi

\IEEEoverridecommandlockouts
 
\usepackage{cite}
\usepackage{amsmath,amssymb,amsfonts}
\usepackage{graphicx}
\usepackage{textcomp}
\usepackage{xcolor}

\usepackage{algorithm}
\usepackage{algorithmic}
\usepackage{subcaption}
\usepackage{verbatim}
\usepackage{listings}

\usepackage{tabularx}
\usepackage{multirow}
\usepackage{pifont}

\usepackage{etoolbox}
\makeatletter
\patchcmd{\@makecaption}
  {\scshape}
  {}
  {}
  {}
\makeatother

\usepackage[]{caption}
\usepackage[strings]{underscore}

\def\BibTeX{{\rm B\kern-.05em{\sc i\kern-.025em b}\kern-.08em
    T\kern-.1667em\lower.7ex\hbox{E}\kern-.125emX}}

\begin{document}

\title{Provisioning Spot Instances Without Employing Fault-Tolerance Mechanisms}

\author{\IEEEauthorblockN{Abdullah Alourani}
\IEEEauthorblockA{University of Illinois at Chicago\\
Chicago, IL 60607\\
aalour2@uic.edu}
\and
\IEEEauthorblockN{Ajay D. Kshemkalyani}
\IEEEauthorblockA{University of Illinois at Chicago\\
Chicago, IL 60607\\
ajay@uic.edu}
}

\maketitle

\begin{abstract}
Cloud computing offers a variable-cost payment scheme that allows cloud customers to specify the price they are willing to pay for renting spot instances to run their applications at much lower costs than fixed payment schemes, and depending on the varying demand from cloud customers, cloud platforms could revoke spot instances at any time. To alleviate the effect of spot instance revocations, applications often employ different fault-tolerance mechanisms to minimize or even eliminate the lost work for each spot instance revocation. However, these fault-tolerance mechanisms incur additional overhead related to application completion time and deployment cost. We propose a novel cloud market-based approach that leverages cloud spot market features to provision spot instances without employing fault-tolerance mechanisms to reduce the deployment cost and completion time of applications. We evaluate our approach in simulations and use Amazon spot instances that contain jobs in Docker containers and realistic price traces from EC2 markets. Our simulation results show that our approach reduces the deployment cost and completion time compared to approaches based on fault-tolerance mechanisms.

\end{abstract}

\begin{IEEEkeywords}

cloud computing; spot instances; fault-tolerance mechanisms; cloud spot market features; cloud-based applications; payment schemes; spot instance revocations

\end{IEEEkeywords}

\IEEEpeerreviewmaketitle

\section{Introduction}\label{II}

Cloud computing offers a variable-cost payment scheme that allows cloud customers to specify the price they are willing to pay for renting spot instances to run their applications at much lower costs than fixed payment schemes, and depending on the varying demand from cloud customers, cloud platforms could revoke spot instances at any time. The price of a spot instance can go up if the demand increases and the number of available instances that can be supported by a finite number of physical resources in a data center of cloud providers decreases. Conversely, the price of this spot instance can go down if the demand decreases and the number of available instances increases. Therefore, if the customer's price is greater than the cloud provider's price that depends on the demand, a spot instance will be provisioned to cloud customers' applications at the customer's price. However, when spot instances are already provisioned to cloud customer applications and the cloud provider's price goes above the customer's price, the cloud providers will terminate those spot instances within two minutes by sending termination notification signals \cite{DBLP:conf/IEEEcloud/AlouraniKG19}. As a result, even though cloud customers sometimes rent spot instances at 90\% lower prices than on-demand prices \cite{SpotPricing}, their applications that run on spot instances can be terminated based on price fluctuations that happen frequently; thus, those applications may incur additional overhead related to application completion time and deployment cost from re-executing lost work for each spot instance revocation.

Applications may benefit from different fault-tolerance mechanisms to alleviate the work lost for each spot instance revocation. However, these fault-tolerance mechanisms incur additional overhead related to application completion time and deployment cost. Fault-tolerance mechanisms are typically divided into three types: migration, checkpointing, and replication. First, migration mechanisms are often employed to reactively migrate the state of an application (i.e., memory and local disk state) to another instance prior to a spot instance revocation. The overhead of a migration mechanism is determined based on the migration time of an application and the number of spot instance revocations during the application execution. The migration time of an application mostly depends on the resource usage of the application, whereas the number of spot instance revocations depends on the volatility of cloud spot markets. A larger resource usage of an application often results in a higher overhead of a migration mechanism, and vice versa. A similar explanation is applicable for the volatility of cloud spot markets; thus, a higher overhead of a migration mechanism will lead to a higher overhead of an application's completion time and deployment cost. Second, checkpointing mechanisms are often employed to proactively checkpoint an application's state to remote storage (e.g., AWS S3). The overhead of a checkpointing mechanism is specified based on the time to checkpoint an application's state and the number of checkpoints, which represents how often an application's state is stored in remote storage during the application execution, along with the time to re-execute the lost work from the last checkpoint for each spot instance revocation. The checkpointing time of an application relies on the resource usage of the application and the number of checkpoints typically specified by engineers who maintain applications deployed on spot instances. If engineers specify a large number of checkpoints, the overhead time to re-execute the lost work from the last checkpoint for each spot instance revocation will likely decrease, whereas the overhead time to checkpoint the state of an application will likely increase. Conversely, if engineers specify a small number of checkpoints, the overhead time to checkpoint the state of an application will likely decrease, whereas the overhead time to re-execute the lost work from the last checkpoint for each spot instance revocation will likely increase. Hence, checkpointing mechanisms require analyzing cloud spot markets and the resource usage of applications to optimize the tradeoff between the overhead of actual checkpoints and the overhead of re-executing lost work. Third, replication mechanisms are often employed to replicate the computations of an application among different instances. The overhead of a replication mechanism is based on the degree of replication (i.e., the number of replicated instances) and the number of revocations that depends on the volatility of cloud spot markets, and is independent of the resource usage of an application. As a result, a higher overhead of these fault-tolerance mechanisms leads to a higher overhead related to application completion time and deployment cost.

\textbf{Contributions:} We address a challenging problem for applications deployed on cloud spot instances that results from the overhead of employing fault-tolerance mechanisms. We propose a novel cloud market-based approach that leverages features of cloud spot markets for provisioning spot instances without employing fault-tolerance mechanisms (\texttt{P-SIWOFT}) to reduce the deployment cost and completion time of applications. We develop \texttt{P-SIWOFT} based on cloud spot market features, such as the spot instance lifetime, revocation probability, and revocation correlation between cloud spot markets and provision spot instances, without employing fault-tolerance mechanisms. We evaluate \texttt{P-SIWOFT} in simulations and use Amazon spot instances that contain jobs in Docker containers and realistic price traces from EC2 markets. Our simulation results show that our approach reduces the deployment cost and completion time compared to approaches based on fault-tolerance mechanisms. \texttt{P-SIWOFT} code and our simulation results are publicly available \cite{P-SIWOFT}.

\section{Problem Statement}\label{PS}
In this section, we discuss sources of overhead of fault-tolerance mechanisms and formulate the problem statement.

\subsection{Sources of Overhead of Fault-Tolerance Mechanisms}

There are three main sources of overhead of fault-tolerance mechanisms. First, various resource usage of an application imposes various overhead of fault-tolerance mechanisms depending on the settings of each fault-tolerance mechanism type. A larger resource usage of an application (i.e., memory footprint) often results in a higher overhead of a fault-tolerance mechanism, and vice versa. The time to migrate/checkpoint the state of an application depends on the sizes of the application's memory and local disk state. Additionally, the choice of the type of fault-tolerance mechanism depends on the resource usage of an application. For example, a live migration requires a limited size of an application's memory footprint and cannot be employed when the application's memory footprint is greater than 4 GB \cite{subramanya2015spoton}. As a result, the resource usage of an application not only affects the overhead of a fault-tolerance mechanism but also affects the choice of the type of fault-tolerance mechanism.

Second, the volatility of cloud markets is represented by the number of spot instance revocations over the application runtime. A higher number of spot instance revocations often results in higher overhead of fault-tolerance mechanisms, and vice versa. Checkpointing mechanisms will re-execute the lost work from the last checkpoint for each spot instance revocation, whereas migration mechanisms will reactively migrate an application to another instance prior to each spot instance revocation. Unlike migration and checkpointing mechanisms, a replication mechanism might re-execute the lost work from the beginning of an application's runtime for each spot instance revocation when all replicated instances are being revoked. As a result, the volatility of cloud markets has an impact on the overhead of various types of fault-tolerance mechanisms.

Third, the overhead of fault-tolerance mechanisms relies on the settings of each type of fault-tolerance mechanism. A main parameter of replication settings is the degree of replication, which represents the number of replicated servers needed to execute the same application's job across these replicated servers. When the degree of replication is small, the overhead that results from re-executing the lost work from the beginning of an application's runtime for each spot instance revocation will likely increase. In contrast, when the degree of replication is large, the overhead that results from a high number of servers will likely increase. A main parameter of checkpointing settings is the number of checkpoints, which represents how often an application's state is stored in remote storage over the application runtime. When the number of checkpoints is small, the overhead that results from re-executing the lost work from the last checkpoint for each spot instance revocation will likely increase. In contrast, when the number of checkpoints is large, the overhead that results from the time to checkpoint an application's state will likely increase. A main parameter of migration settings is the number of migrations, which represents how often an application's state migrates to another server over the application runtime. When the number of migrations is small, the overhead that results from re-executing the lost work from the beginning of an application's runtime for each spot instance revocation will likely increase. In contrast, when the number of migrations is large, the overhead that results from the time to migrate an application's state will likely increase. As a result, the fundamental problem for cloud customers is determining how to find the optimal settings of various types of fault-tolerance mechanisms to reduce the overhead resulting from employing fault-tolerance mechanisms.

\subsection{The Problem Statement}

Cloud computing offers a variable-cost payment scheme that allows cloud customers to specify the price they are willing to pay for renting spot instances to run their applications at much lower costs than fixed payment schemes. In exchange, applications deployed on spot instances are often exposed to revocations by cloud providers, and as a result, these applications often employ different fault-tolerance mechanisms to minimize or even eliminate the lost work for each spot instance revocation. However, the overhead resulting from employing fault-tolerance mechanisms has become a very important concern for cloud customers. In this paper, we address a challenging problem for applications deployed on cloud spot instances that results from the overhead of employing fault-tolerance mechanisms—determining how to effectively deploy applications on spot instances without employing fault-tolerance mechanisms to reduce the deployment cost and completion time of applications. The root of this problem is that applications often employ fault-tolerance mechanisms to minimize the lost work for each spot instance revocation without taking into consideration the overhead of fault-tolerance mechanisms, leading to significantly larger deployment costs and completion times of applications, and as a result, the advantages of cloud spot instances could be significantly minimized or even completely eliminated.

\section{Our Approach}\label{SS}

In this section, we state our key ideas for \texttt{P-SIWOFT} and explain the \texttt{P-SIWOFT} algorithm.

\subsection{Key Ideas}\label{Ideas}

A goal of our approach is to automatically provision spot instances without employing fault-tolerance mechanisms to reduce the deployment cost and completion time of applications. Our approach leverages features of cloud spot markets such as the spot instance lifetime, revocation probability, and revocation correlation between cloud spot markets to provision spot instances for applications. The spot instance lifetime represents the average time until a spot instance's price rises above the corresponding on-demand instance price (i.e., mean time to revocation (MTTR)) because cloud customers are often not willing to pay more than the on-demand price to rent spot instances. The revocation probability of each spot instance represents the estimated lifetime of a spot instance during a job execution and is calculated by dividing the job's execution length by the MTTR of the provisioned spot instance. The revocation correlation between cloud spot instances represents how often these spot instances were revoked at the same time (i.e., the same hour representing a single billing cycle in cloud platforms \cite{SpotPricing}) over the past three months.

In general, cloud spot markets show a broad range of characteristics. These important characteristics are at the core of our approach. First, revocations rarely happen in some cloud spot markets, so the MTTR of these markets is very high (i.e., $>$ 600 h) \cite{196312}. Second, employing fault-tolerance mechanisms often results in additional overhead related to application completion time and deployment cost \cite{subramanya2015spoton}. Third, cloud spot markets exhibit variations in price characteristics for a similar type of spot instance across various cloud spot markets. Thus, a spot instance in a cloud market is often independent of a spot instance in another cloud market, which suggests that a spot instance's revocation in a cloud market is often uncorrelated with a spot instance in another cloud market \cite{sharma2017portfolio}. Based on these characteristics, our key idea is that we could eliminate the additional overhead resulting from employing fault-tolerance mechanisms by provisioning the spot instance with the highest MTTR as long as the spot instance's MTTR is at least twice the application's execution length. Another idea is that we could reduce consequent revocations when a spot instance is revoked by provisioning a new spot instance with the next highest MTTR and a low revocation correlation with the revoked spot instance. When we provision a spot instance that is uncorrelated with the revoked spot instance, it is more unlikely that the new spot instance will be revoked again than another spot instance that is highly correlated with the revoked spot instance. As a result, these key ideas enable cloud customers to avoid unnecessary overhead resulting from employing fault-tolerance mechanisms; hence, cloud customers can execute jobs with a completion time near that of on-demand instances but at a cost of only spot instances.

\setlength{\textfloatsep}{4pt}
\begin{algorithm}[t]  
\caption{\texttt{P-SIWOFT}'s algorithm for provisioning spot instances without employing fault-tolerance mechanisms.}
\label{alg:MOPTICLE}
\begin{algorithmic}[1]

\STATE \textbf{\texttt{Inputs:}} Jobs $J$, Cloud Markets $M$, Resources $R$

\STATE $U \leftarrow$ \textbf{\texttt{FindSuitableServers}}($ J$, $ R$) 
\STATE $ L \leftarrow$ \textbf{\texttt{ComputeLifeTime}}($ M$, $ U$) 

\FOR {each j \textbf{in} $J$ }

\STATE $S_{j}$ $\leftarrow$ \textbf{\texttt{ServerBasedLifeTime}}($j$, $M$, $L$)

\WHILE {$j$ $\neg$ \textbf{\texttt{Completed}} }

\STATE $s_{j}$ $\leftarrow$ \textbf{\texttt{Highest}}($S_{j}$)

\IF{ $length(s_j)$ $>>$ $length(j)$}

\STATE $v_{s_j}$ $\leftarrow$ \textbf{\texttt{RevocationProbability}}($j$, $s_{j}$)
\STATE \textbf{\texttt{ProvisionHighestLifeTime}}($j$, $s_{j}$)

\IF{ $s_{j}$ $encounters$ $v_{s_j}$}

\STATE $C_{j}$, $T_{j}$ $\leftarrow$
$C_{j}$ $\cup$ \{$c_{s_{j}}$\}, $T_{j}$ $\cup$ \{$t_{s_{j}}$\}

\STATE $W_{s_{j}}$ $\leftarrow$ \textbf{\texttt{FindLowCorrelation}}($j$, $s_{j}$))

\STATE $S_{j}$ $\leftarrow$ ($S_{j}$ $\backslash$ \{$s_{j}$\}) $\cap$  $W_{s_{j}}$

\ENDIF

\ENDIF

\ENDWHILE

\STATE $C_{j}$, $T_{j}$ $\leftarrow$
$C_{j}$ $\cup$ \{$c_{s_{j}}$\}, $T_{j}$ $\cup$ \{$t_{s_{j}}$\}

\STATE $C, T \leftarrow$ \textbf{\texttt{ComputeCostExeTime}}($C_{j}$, $T_{j}$)

\ENDFOR

\RETURN {${C}$, ${T}$}
\end{algorithmic}
\end{algorithm}

\subsection{P-SIWOFT Algorithm}

\texttt{P-SIWOFT} is illustrated in Algorithm \ref{alg:MOPTICLE} that takes in the batch job set $J$; the resource requirement set $R$; and the entire set of cloud markets $M$, containing on-demand instance types, prices of on-demand instances, spot instance types, their availability zones, their regions, and spot instance prices over the past three months. Starting from Step 2, the algorithm finds a suitable set of spot instances $U$ that meet the resource requirements. In \texttt{P-SIWOFT}, we use the memory size to determine suitable types of spot instances that are supported by EC2 markets \cite{SpotPricing}. In Step 3, for each suitable spot instance, the spot instance’s lifetime (i.e., the spot instance’s MTTR) is computed based on the corresponding on-demand instance price, as discussed in Section \ref{Ideas}. $L$ is the set of such lifetimes. \textcolor{black}{In Steps 4-20, for each job, the algorithm is executed until the jobs in the job set are completed.} In Step 5, the cloud spot markets are first filtered to include only a set of suitable spot instances $S_{j}$ for the job $j$ according to their lifetimes $L$, as discussed in Section \ref{Ideas}, and then these spot instances are sorted in descending order based on their lifetimes. In Steps 6–17, the job $j$ is executed until the job’s execution is completed. In Step 7, the algorithm selects a spot instance $s_{j}$ with the highest lifetime. In Step 8, we ensure that the highest lifetime for the spot instance $s_{j}$ is at least twice the job $j's$ execution length to reduce the revocation probability of the provisioned instance during the job execution. In Step 9, the algorithm computes the revocation probability of the provisioned instance $v_{s_j}$ by dividing the job $j's$ execution length by the lifetime of the provisioned instance $s_{j}$. In Step 10, the spot instance $s_{j}$ with the highest lifetime is provisioned to (re)start executing the job $j$. In Steps 11–15, the algorithm checks if the provisioned spot instance $s_{j}$ is revoked based on its revocation probability $v_{s_j} $ during the job execution $j$. \textcolor{black}{When a spot instance $s_{j}$ is revoked, the deployment time $t_{s_{j}}$ and cost $c_{s_{j}}$ are added to the total deployment time set $T_{j}$ and cost set $C_{j}$, respectively, in Step 12.} In \texttt{P-SIWOFT}, the deployment time represents the job’s execution time until the spot instance is revoked, the deployment cost of a spot instance represents the price of the provisioned spot instance at a certain execution point, and the cost is computed at a per hour rate (i.e., a single billing cycle in cloud platforms \cite{SpotPricing}). In Step 13, the low revocation correlation set $W_{s_{j}}$ with the revoked spot instance is computed using the revocation correlation between cloud spot instances, as discussed in Section \ref{Ideas}. In Step 14, the revoked spot instance is removed from the set of suitable spot instances $S_{j}$, and \textcolor{black}{the set of suitable spot instances $S_{j}$ is filtered based on a low revocation correlation set $W_{s_{j}}$.} The cycle of Steps 6–17 repeats until the job $j's$ execution is completed. \textcolor{black}{When the job $j's$ execution is completed, the deployment time $t_{s_{j}}$ and cost $c_{s_{j}}$ are added to the total deployment time set $T_{j}$ and cost set $C_{j}$, respectively, in Step 18. In Step 19, the total deployment time set $T_{j}$ and cost set $C_{j}$ are computed and then added to the overall deployment time $T$ and cost $C$, respectively. The cycle of Steps 4–20 repeats until the jobs in the job set are completed.} Finally, the total deployment time $T$ and cost $C$ are returned in Step 21 as the algorithm ends.

\section{Empirical Evaluation}\label{EE}

In this section, we describe the design of the empirical study to evaluate
\texttt{P-SIWOFT} and state threats to its validity. We pose the following Research Questions (RQs):

\begin{description}

 \item[\textbf{\textit{RQ}}$_{1}$:] How efficient is \texttt{P-SIWOFT} compared to a fault-tolerance approach in executing applications?
 \item[\textbf{\textit{RQ}}$_{2}$:] How effective is \texttt{P-SIWOFT} compared to a fault-tolerance approach in reducing the deployment cost of applications?
\item[\textbf{\textit{RQ}}$_{3}$:] Do different settings of a fault-tolerance approach contribute to different types of overhead?

\end{description}

\subsection{Subject applications}

We evaluate \texttt{P-SIWOFT} in simulations and use Amazon spot instances that contain jobs in Docker containers and realistic price traces from EC2 markets. \texttt{P-SIWOFT} packages jobs in Docker containers to simplify restoring and checkpointing. We use a load generator called Lookbusy \cite{lookbusy:2020} to create synthetic jobs with different amounts of resource usage. In addition, \texttt{P-SIWOFT} uses EC2's REST API to collect realistic price traces for all spot instances across all markets (i.e., availability zones and regions) for the past three months.

\subsection{Methodology}

Some objectives of the experiments are to demonstrate that \texttt{P-SIWOFT} can efficiently execute applications and can effectively decrease the deployment cost of applications compared to a fault-tolerance approach. For these objectives, we use different combinations of job execution length and job memory footprint to show the impact on the completion time and the deployment cost when a spot instance is provisioned for the job using \texttt{P-SIWOFT} and the fault-tolerance approach. We define two revocation rules with different ranges for \texttt{P-SIWOFT} and the fault-tolerance approach to show the impact on 
the completion time and the deployment cost for different numbers of revocations during a job's execution. When a spot instance is provisioned for a job using the fault-tolerance approach, we randomly send a fixed number of revocations per day of the job's execution length, as suggested by prior work \cite{subramanya2015spoton}. Conversely, when a spot instance is provisioned for a job using \texttt{P-SIWOFT}, we use the revocation probability of a spot instance that relies on realistic price traces from the Amazon cloud to revoke the provisioned spot instance. Since another goal is to understand how different settings of jobs and different settings of the fault-tolerance approach contribute to different types of overhead (e.g., checkpoint overhead), we investigate how different job execution lengths, job memory footprints, numbers of revocations, and numbers of checkpoints contribute to different overhead types that are related to a job's completion time and deployment cost.

\texttt{P-SIWOFT} is implemented using a load generator API (Lookbusy), EC2's REST API, Docker containers, AWS S3, and EC2 spot instances. The experiments for the subject applications were carried out using spot instances from Amazon EC2 called m5ad.12xlarge with a 48 GHz CPU and 192 GB of memory. We package jobs in Docker containers that run on Ubuntu 18.04 LTS with a limited CPU and memory capacity for the provisioned spot instances to assess the effectiveness of \texttt{P-SIWOFT} for different job memory footprints and job execution lengths.

\subsection{Threads to validity }

One potential threat to our empirical evaluation is that our experiments were conducted only on batch job applications, which may make it difficult to generalize the results of the experiments to other types of applications (e.g., interactive job applications) that may have various workflows and behaviors. However, cloud spot instances are often used to run batch job applications. As a result, we expect the results of the experiments to be generalizable.

We experimented with a certain price ratio between spot instances and on-demand instances that is based on realistic price traces from EC2 markets, whereas other ratios between spot instances and on-demand instances could result in different effects on the deployment cost and completion time of jobs when spot instances are provisioned using \texttt{P-SIWOFT} and the fault-tolerance approach. However, understanding the effect of various price ratios between spot instances and on-demand instances is beyond the scope of this empirical study and shall be considered in future studies.

\section{EMPIRICAL RESULTS}

In this section, we describe and analyze the results of the experiments to
answer the\ RQs listed in Section~\ref{EE}.

\begin{figure*}[t!] 
\begin{subfigure}{0.35\textwidth}\centering
\includegraphics[width=0.8\textwidth]{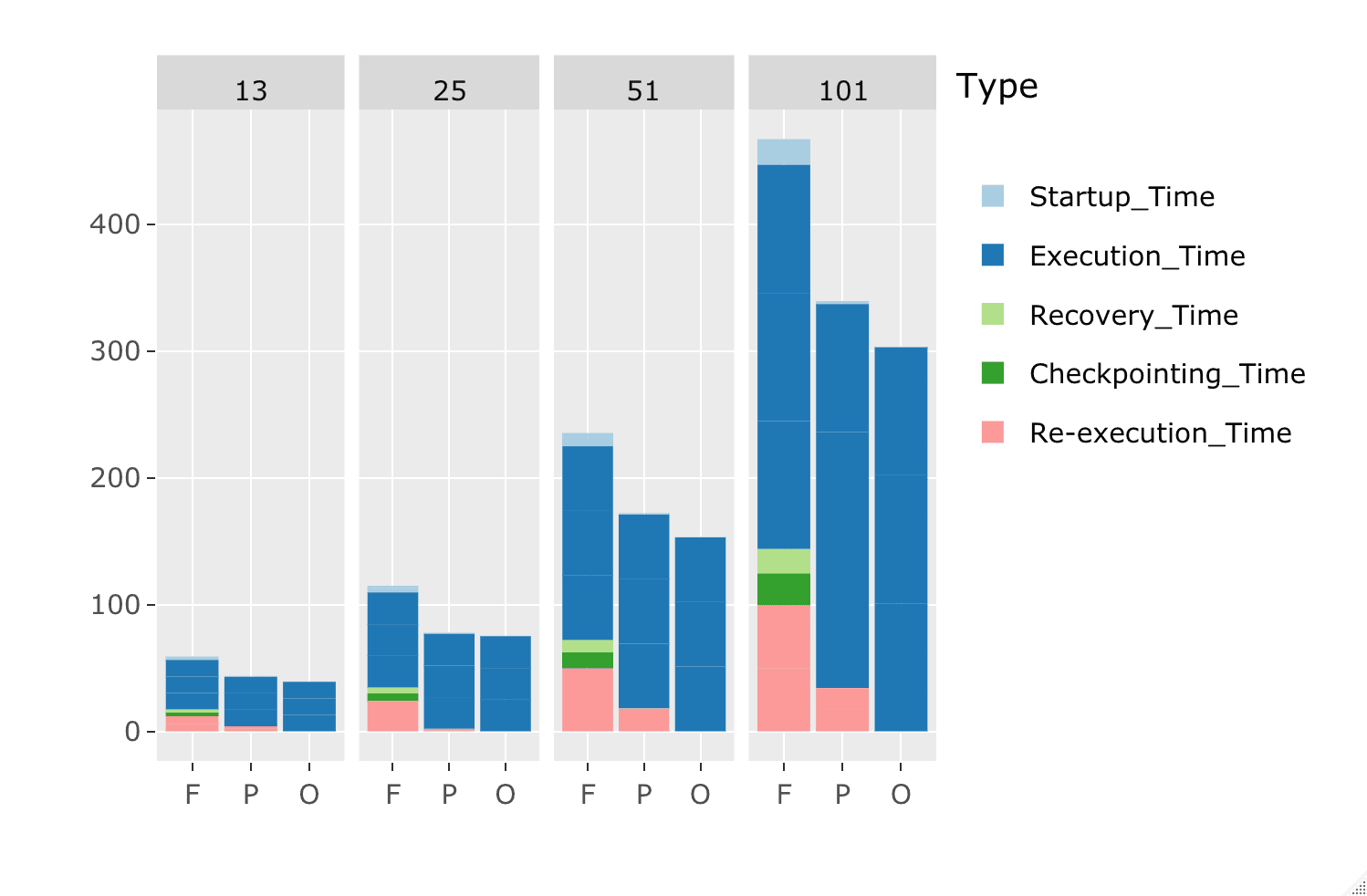}
\vspace{-0.3cm}
\caption{Job Length (Time)} \label{fig:a}
\end{subfigure}\hspace*{\fill}
\begin{subfigure}{0.35\textwidth}\centering
\includegraphics[width=0.8\linewidth]{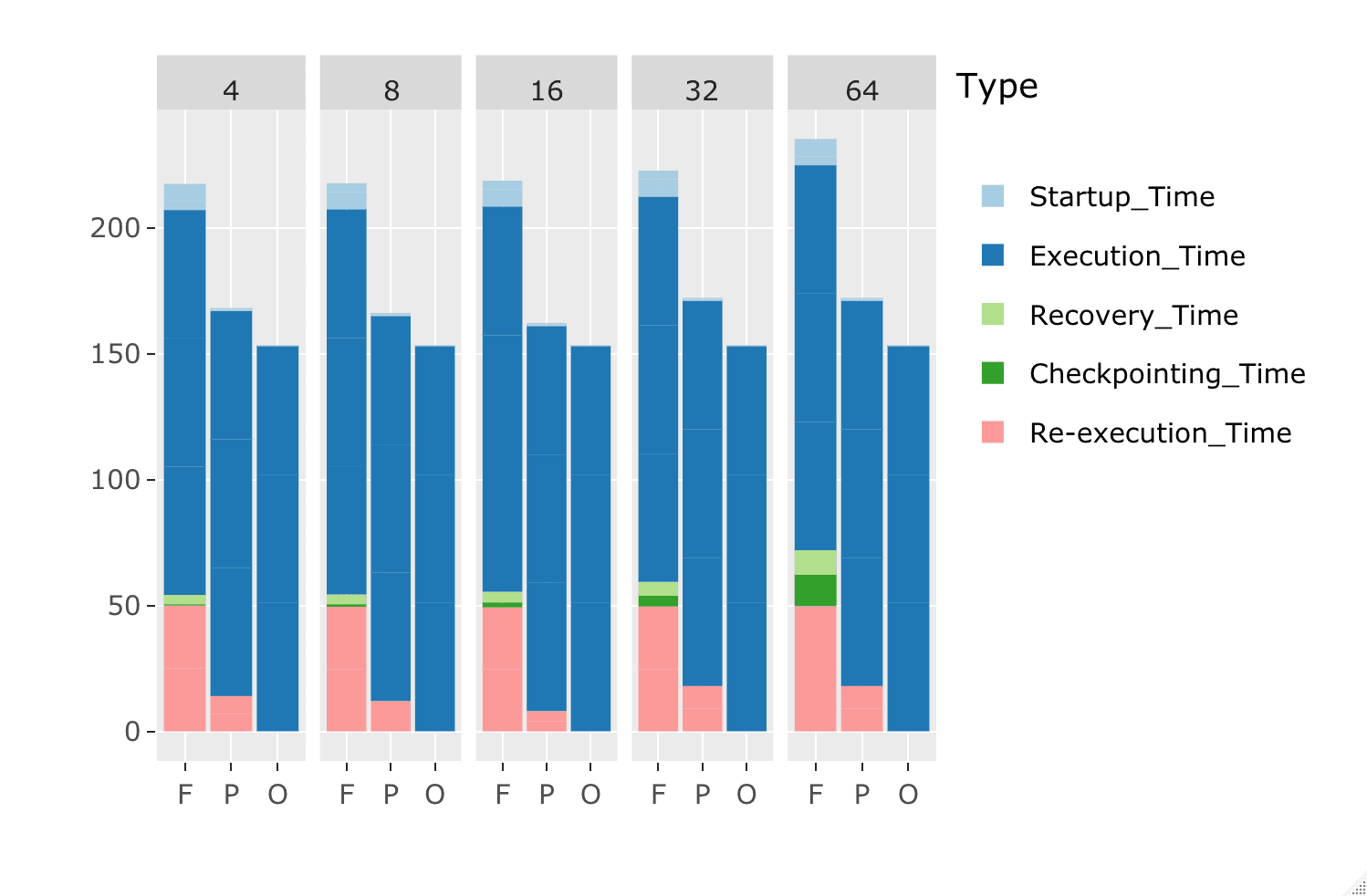}
\vspace{-0.3cm}
\caption{Memory Footprint (Time)} \label{fig:b}
\end{subfigure}\hspace*{\fill}
\begin{subfigure}{0.35\textwidth}\centering
\includegraphics[width=0.8\linewidth]{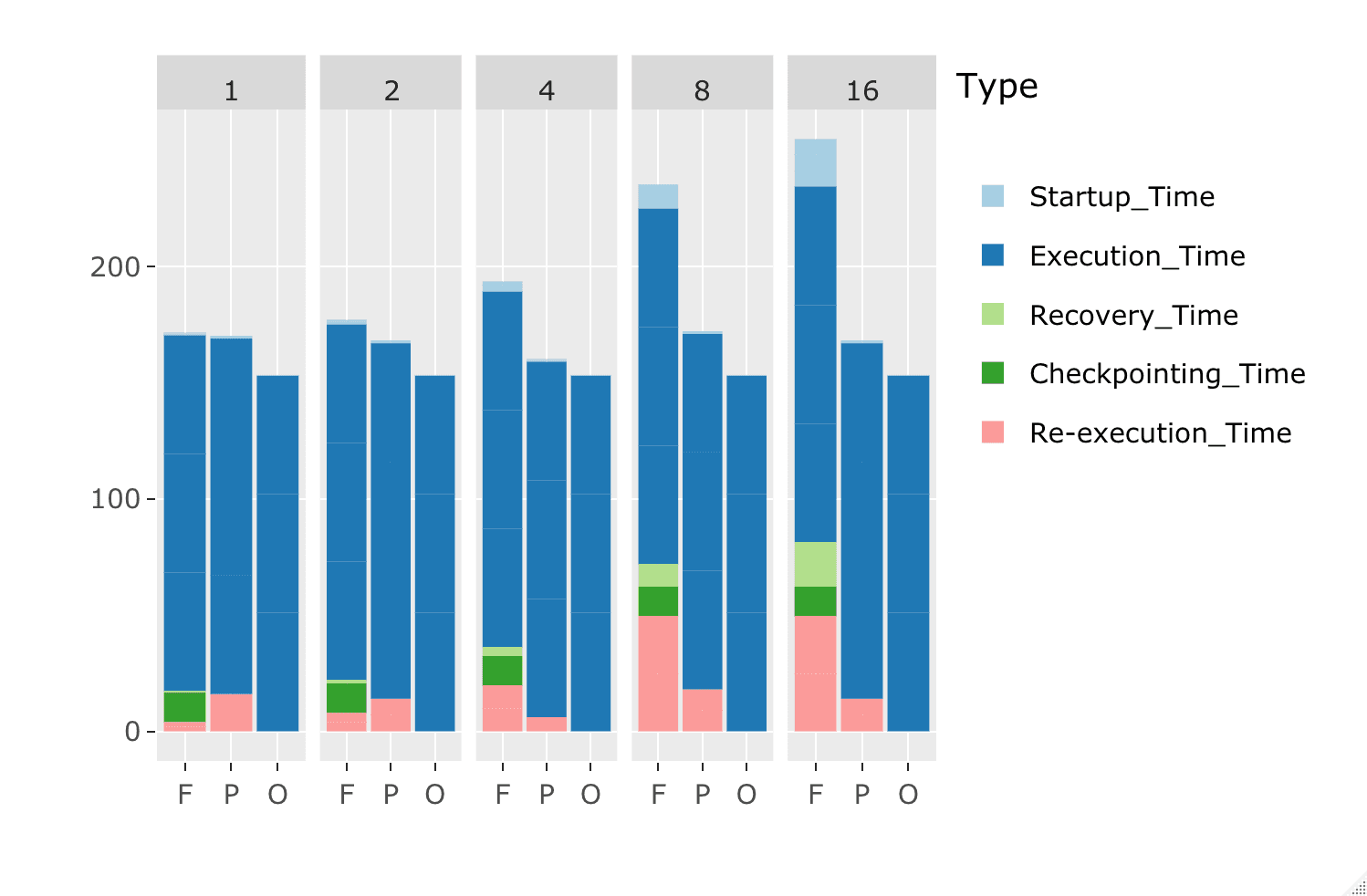}
\vspace{-0.3cm}
\caption{Revocation Number (Time)} \label{fig:c}
\end{subfigure}
\vspace{-0.3cm}
 
\medskip
 \begin{subfigure}{0.35\textwidth}\centering

\includegraphics[width=0.8\linewidth]{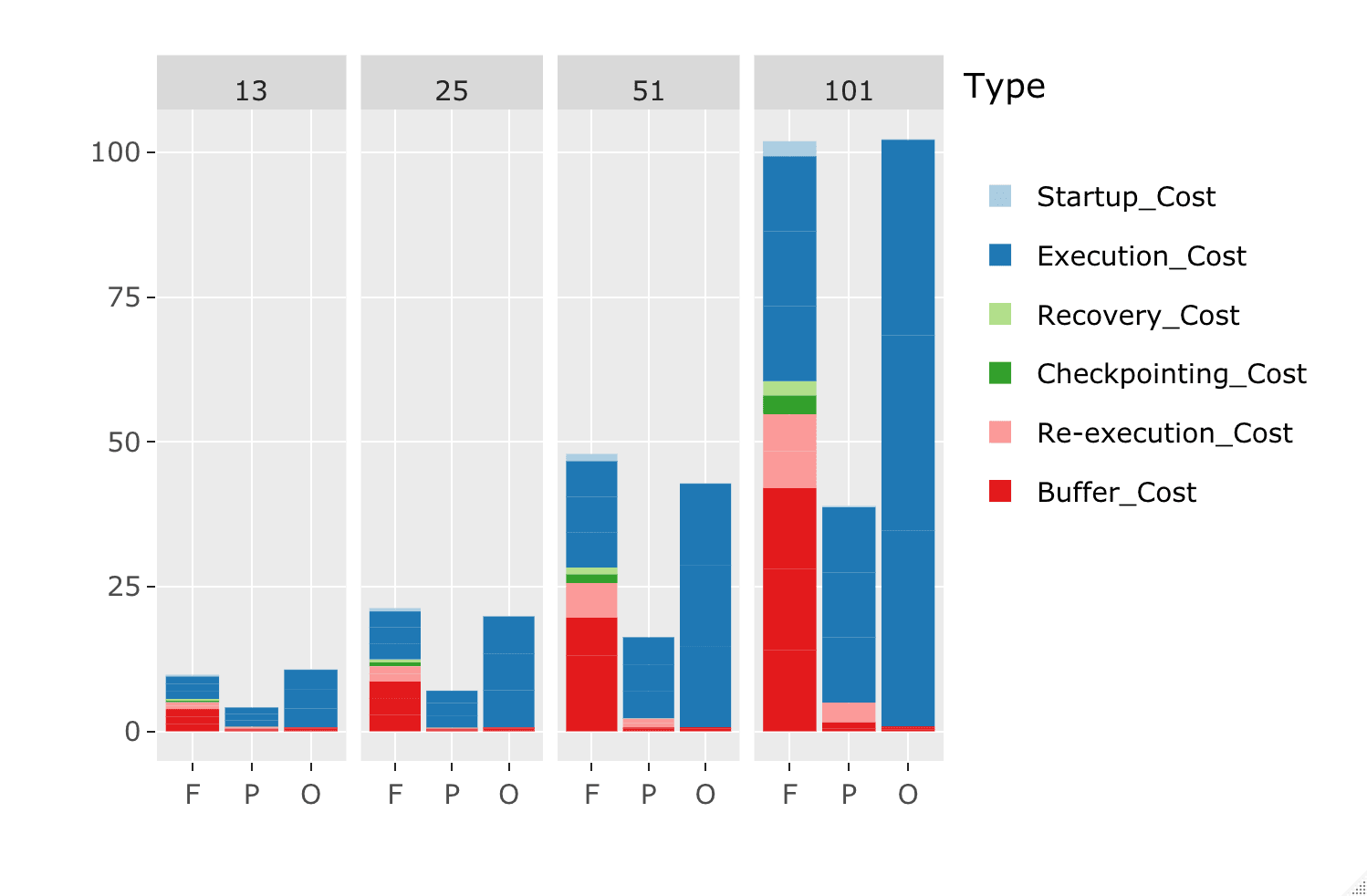}

\vspace{-0.3cm}
\caption{Job Length (Cost)} \label{fig:d}
\end{subfigure}\hspace*{\fill}
\begin{subfigure}{0.35\textwidth}\centering
\includegraphics[width=0.8\linewidth]{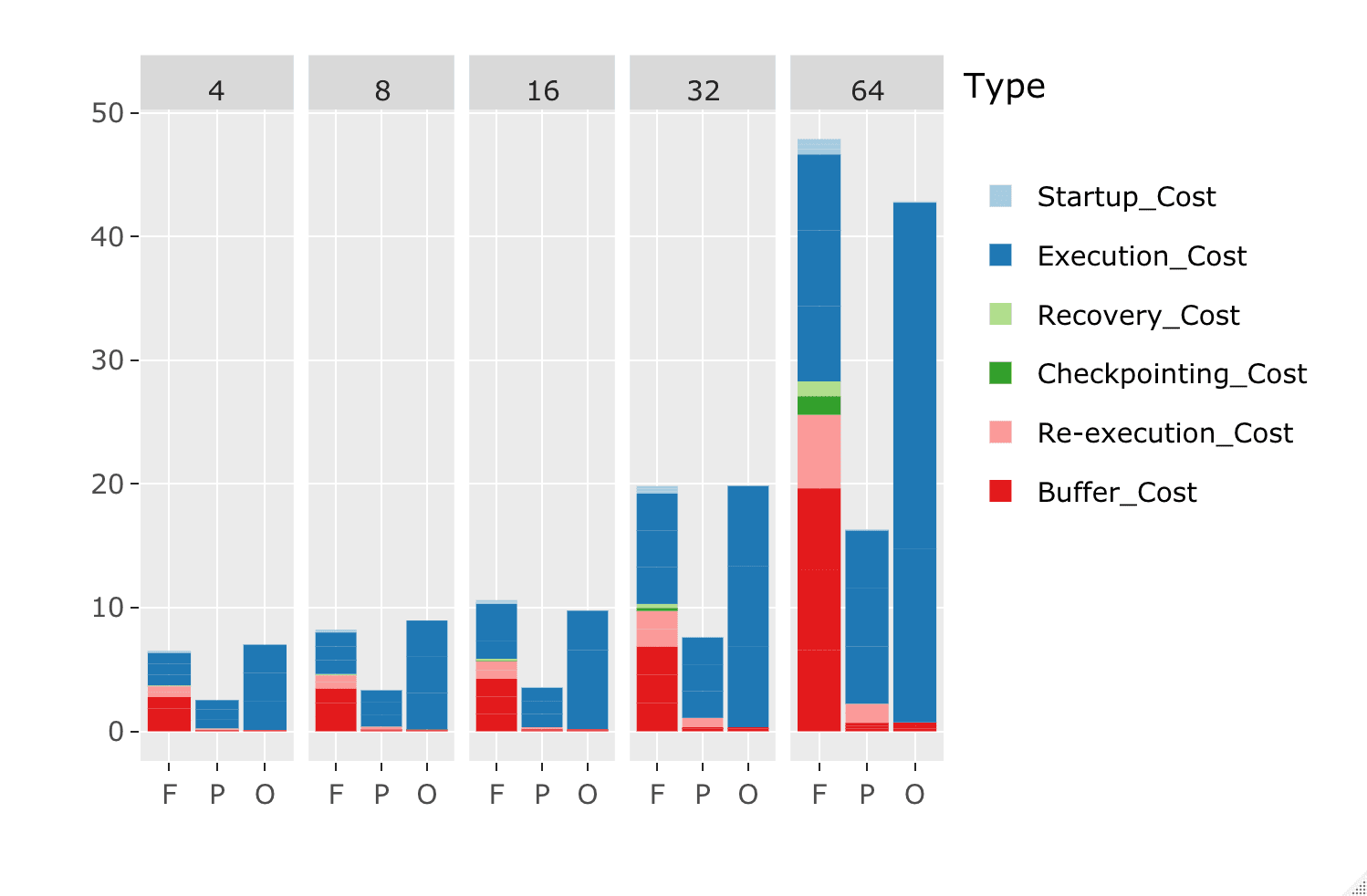}
 
\vspace{-0.3cm}
\caption{Memory Footprint (Cost)} \label{fig:e}
 \end{subfigure}\hspace*{\fill}
\begin{subfigure}{0.35\textwidth}\centering
\includegraphics[width=0.8\linewidth]{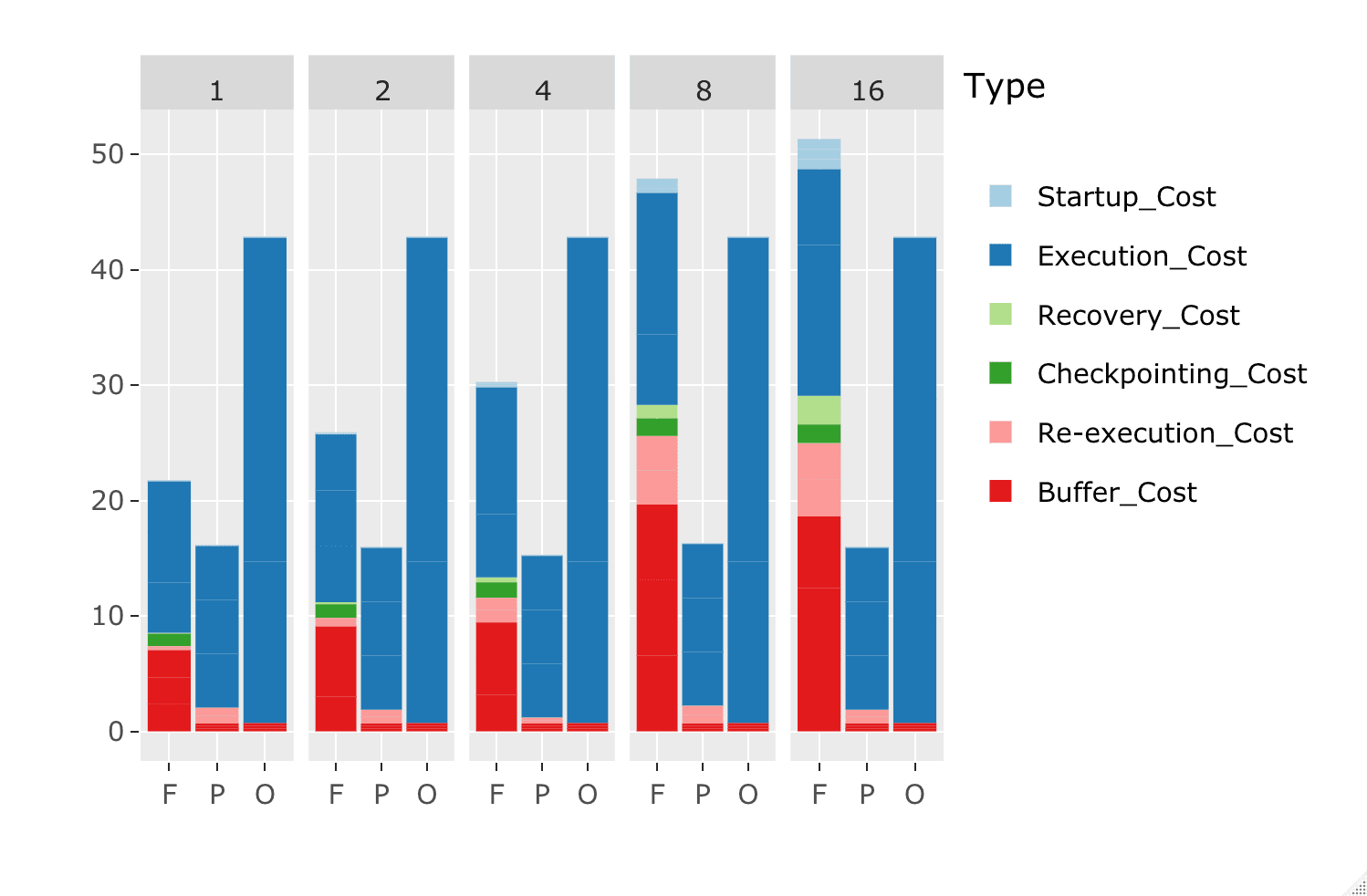}
\vspace{-0.3cm}
 \caption{Revocation Number (Cost)} \label{fig:f}
\end{subfigure}
 \caption{Comparing the completion time (top row) and the deployment costs (bottom row) for the subject applications using \texttt{P-SIWOFT} (P), the fault-tolerance approach (F), and on-demand instances (O) for different job execution lengths (a and d), memory footprints (b and e), and revocation numbers (c and f), while keeping other job features constant.} \label{fig:all}
\vspace{-2.5ex}

\end{figure*}

\subsection{Completion Time }

The experimental results that summarize the completion time for the subject applications using \texttt{P-SIWOFT}, the fault-tolerance approach, and on-demand instances for different job execution lengths are shown in the stacked bar plots in Fig.~\ref{fig:a}. We observe that the completion time using \texttt{P-SIWOFT} is consistently shorter than the completion time using the fault-tolerance approach, and the completion time using \texttt{P-SIWOFT} is consistently near that of on-demand instances, which do not incur any additional overhead \cite{SpotPricing}. This result shows that a higher job length leads to a steadily higher overhead of completion time resulting from the job’s checkpointing, recovery, and re-execution times, as well as the startup time of a spot instance when using the fault-tolerance approach. However, a higher job length leads to a slightly higher overhead of the completion time, as a result of the job’s re-execution time and the startup time of a spot instance when using \texttt{P-SIWOFT}. Our explanation is that \texttt{P-SIWOFT} does not incur frequent job re-execution time and the startup time of a spot instance since the startup time of a spot instance using \texttt{P-SIWOFT} does not increase with the increase in job execution length. This is expected based on the way \texttt{P-SIWOFT} provisions a spot instance with the highest MTTR. 

The experimental results that summarize the completion time for the subject applications using \texttt{P-SIWOFT}, the fault-tolerance approach, and on-demand instances for different job memory footprints are shown in the stacked bar plots in Fig.~\ref{fig:b}. We observe that the completion time for \texttt{P-SIWOFT} is consistently shorter than the completion time for the fault-tolerance approach, and the completion time for \texttt{P-SIWOFT} is consistently near that of on-demand instances, which do not incur any additional overhead \cite{SpotPricing}. This result shows that a higher job memory footprint leads to a higher overhead of the completion time resulting from the job’s checkpointing time and recovery time when using the fault-tolerance approach. In contrast, the overhead of the completion time resulting from the job’s re-execution time and the startup time of a spot instance when using the fault-tolerance approach stays approximately the same across various job memory footprints, which suggests that the overhead resulting from the job’s re-execution time and the startup time of a spot instance for the fault-tolerance approach is independent of the job resource usage. Also, the overhead of an application’s completion time resulting from the job’s re-execution time and the startup time of a spot instance when using \texttt{P-SIWOFT} stays approximately the same across various job memory footprints, which suggests that the completion time for the subject applications when using \texttt{P-SIWOFT} is also independent of the resource usage. 

The experimental results that summarize the completion time for the subject applications using \texttt{P-SIWOFT}, the fault-tolerance approach, and on-demand instances for different numbers of revocations are shown in the stacked bar plots in Fig.~\ref{fig:c}. We observe that the completion time for \texttt{P-SIWOFT}—except for when the number of revocations equals one—is consistently shorter than the completion time for the fault-tolerance approach, and the completion time for \texttt{P-SIWOFT} is consistently near that of on-demand instances, which do not incur any additional overhead \cite{SpotPricing}. When the number of revocations equals one, the job’s checkpointing time for the fault-tolerance approach balances the job’s re-execution for \texttt{P-SIWOFT}. This result suggests that the fault-tolerance approach incurs additional overhead due not only to the number of revocations, but also the number of checkpoints. It also suggests that the effectiveness of \texttt{P-SIWOFT} may decrease when the number of revocations decreases, and it is very difficult to guarantee that the number of revocations is small \cite{shastri2017hotspot}. The job’s recovery time, the job’s re-execution time, and the startup time of a spot instance—except for the job’s checkpointing time—all increase steadily when using the fault-tolerance approach, whereas in \texttt{P-SIWOFT}, the job’s re-execution time and the startup time of a spot instance stay approximately the same. This observation suggests that the job’s checkpointing time for the fault-tolerance approach as well as the job’s re-execution time and the startup time of a spot instance for \texttt{P-SIWOFT}, are independent of the number of revocations. In summary, these experimental results allow us to conclude that \texttt{P-SIWOFT} is more efficient in executing applications for different job execution lengths, job memory footprints, and numbers of revocations than the fault-tolerance approach, thus \textbf{positively addressing} \textbf{\textit{RQ}}$_{1}$.

\subsection{Deployment Costs}

The experimental results that summarize the deployment costs for the subject applications using \texttt{P-SIWOFT}, the fault-tolerance approach, and on-demand instances for different job execution lengths are shown in the stacked bar plots in Fig.~\ref{fig:d}. We observe that the deployment costs using \texttt{P-SIWOFT} are consistently lower than the deployment costs using the fault-tolerance approach or those of on-demand instances. This result identifies the steady rise in overhead related to deployment costs that result from the job’s checkpointing costs, its recovery costs, its re-execution costs, the startup costs of spot instances, and the buffer costs of billing cycles when using the fault-tolerance approach with the increased job length. However, this result also identifies a slight rise in the overhead of deployment costs that result from the job’s re-execution cost, the startup costs of spot instances, and the buffer costs of billing cycles when using \texttt{P-SIWOFT} with the increased length. Our explanation is that \texttt{P-SIWOFT} does not frequently incur the job’s re-execution costs and the startup costs of spot instances since the startup costs of spot instances using \texttt{P-SIWOFT} do not increase with the increase of the job execution length, which is expected based on the way that \texttt{P-SIWOFT} provisions a spot instance with the highest MTTR. Interestingly, we observe that unlike \texttt{P-SIWOFT}, the buffer costs of billing cycles significantly increase compared to the other types of overhead costs when using the fault-tolerance approach with the increase of the job length, which suggests that the fault-tolerance approach incurs not only overhead related to the settings of the fault-tolerance approach (e.g., the job’s checkpointing cost) but also additional overhead related to the cloud billing policies (i.e., the buffer costs of billing cycles). Also, we observe that the deployment costs of the fault-tolerance approach across all job lengths are equal to or higher than the deployment costs of on-demand instances \cite{SpotPricing}, which suggests using on-demand for larger job lengths may reduce deployment costs and the completion time when compared to the fault-tolerance approach. 

The experimental results that summarize the deployment costs for the subject applications using \texttt{P-SIWOFT}, the fault-tolerance approach, and on-demand instances for different job memory footprints are shown in the stacked bar plots in Fig.~\ref{fig:e}. We observe that the deployment costs using \texttt{P-SIWOFT} are consistently lower than the deployment costs using the fault-tolerance approach and on-demand instances. This result demonstrates the steady rise of the overhead related to deployment costs resulting from the job’s checkpointing, recovery, re-execution, and startup costs of spot instances, as well as the buffer costs of billing cycles when using the fault-tolerance approach with the increase of job memory footprint. However, this result demonstrates a slight rise of the overhead of deployment costs resulting from the job’s re-execution and startup costs of spot instances, and the buffer costs of billing cycles when using \texttt{P-SIWOFT} with the increase of job memory footprint. Our explanation is that \texttt{P-SIWOFT} does not incur the job’s re-execution and startup costs of spot instances, since the startup costs of spot instances using \texttt{P-SIWOFT} do not increase with the increase of the job memory footprint, which is expected based on the way that \texttt{P-SIWOFT} provisions a spot instance with the highest MTTR. We observe that, unlike the buffer costs of billing cycles for \texttt{P-SIWOFT}, the buffer costs of billing cycles for the fault-tolerance approach significantly increase with the higher job memory footprints (i.e., 32 and 64 GB), suggesting that the buffer costs increase when there is a significant change in deployment time between consecutive job memory footprints (i.e., exceeds the period for a billing cycle). Additionally, we observe that the deployment costs of the fault-tolerance approach across all job memory footprints are equal or higher than the deployment costs of on-demand instances \cite{SpotPricing}, which suggests provisioning on-demand for large job memory footprints may result in lower deployment costs and completion time than the fault-tolerance approach. 

The experimental results that summarize the deployment costs for the subject applications using \texttt{P-SIWOFT}, the fault-tolerance approach, on-demand instances for different numbers of revocations are shown in the stacked bar plots in Fig.~\ref{fig:f}. We observe that the deployment costs using \texttt{P-SIWOFT} and that of on-demand instances are consistently lower than the deployment costs using the fault-tolerance approach. The job’s recovery and re-execution costs, the startup costs of spot instances, and the buffer costs of billing cycles, except for the job’s checkpointing costs, increase steadily when using the fault-tolerance approach whereas, for \texttt{P-SIWOFT}, the job’s re-execution costs, the startup costs of spot instances, and the buffer costs of billing cycles stay approximately the same. This observation suggests that the job’s recovery time and re-execution costs, the startup costs of spot instances, and the buffer costs of billing cycles depend on the number of revocations when using the fault-tolerance approach. However, the job’s checkpointing costs for the fault-tolerance approach and the job’s re-execution costs, the startup costs of spot instances, and the buffer costs of billing cycles for \texttt{P-SIWOFT}, are independent of the number of revocations, respectively. Our explanation is that \texttt{P-SIWOFT} does not incur the job’s re-execution costs and the startup costs of spot instances. We observe that unlike the buffer costs of billing cycles for \texttt{P-SIWOFT}, the buffer costs of billing cycles for the fault-tolerance approach significantly increase with the higher numbers of revocations (i.e., 8 and 16), which suggests that the buffer costs increase when there is a significant change in deployment time between consecutive numbers of revocations (i.e., exceeds the period for a billing cycle). Interestingly, we observe that the deployment costs for the fault-tolerance approach when the number of revocations is high (i.e., 8 and 16) is significantly higher than the deployment costs for on-demand instances \cite{SpotPricing}, which confirms that provisioning on-demand for a large number of revocations may result in lower deployment costs and completion time than the fault-tolerance approach. In summary, these experimental results allow us to conclude that \texttt{P-SIWOFT} is more effective in reducing the deployment costs of applications for different job execution lengths, job memory footprints, and numbers of revocations than the fault-tolerance approach, thus \textbf{positively addressing} \textbf{\textit{RQ}}$_{2}$.

\subsection{Impact on Different Types of Overhead}

An interesting question is how different job execution lengths, job memory footprints, and numbers of revocations, contribute to different overhead types that are related to a job’s completion time and deployment cost when using the fault-tolerance approach. Consider the stacked bar plots that are shown in Fig.~\ref{fig:a}, Fig.~\ref{fig:b}, and Fig.~\ref{fig:c} — the visual inspection identifies the highest overhead related to the completion time results from the job’s re-execution time, then the job’s checkpointing time and the job’s recovery time, followed by the startup time of a spot instance, with the increase of the job execution length. Also, with the rise of the job memory footprint, the highest overhead related to the completion time when using the fault-tolerance approach results from the job’s checkpointing time and the job’s recovery time. With the increase of the number of revocations, the highest overhead related to the completion time when using the fault-tolerance approach results from the job’s re-execution time, then the job’s recovery time, followed by the startup time of a spot instance.

Similarly, it is shown in the stacked bar plots in Fig.~\ref{fig:d}, Fig.~\ref{fig:e}, and Fig.~\ref{fig:f} that the highest overhead related to the deployment costs when using the fault-tolerance approach results from the buffer costs of billing cycles, the job’s re-execution costs, then the job’s checkpointing cost, the job’s recovery cost, followed by the startup costs of spot instances, with the increase of the job execution length. With the rise of the job memory footprint, the highest overhead related to the deployment costs when using the fault-tolerance approach results from the buffer costs of billing cycles, the job’s re-execution costs, then the job’s checkpointing and recovery costs, followed by the startup costs of spot instances. With the increase of the number of revocations, the highest overhead related to the deployment costs when using the fault-tolerance approach results from the buffer costs of billing cycles, the job’s re-execution costs, then its recovery costs, followed by the startup costs of spot instances. The results confirm that different job execution lengths, job memory footprints, and numbers of revocations contribute to different overhead types related to a job’s completion time and deployment cost when using the fault-tolerance approach, thus \textbf{positively addressing} \textbf{\textit{RQ}}$_{3}$.

\section{Related Work}

Most of the prior works focused on reducing the effect of spot instance revocations using fault-tolerance methods, such as replication \cite{voorsluys2012reliable,harlap2017proteus,DBLP:conf/eurosys/SharmaLGIS15}, checkpointing \cite{yi2010reducing,subramanya2015spoton,sharma2017portfolio}, and VM migration \cite{jia2016smart,shastri2017hotspot}. Voorsluys et al. \cite{voorsluys2012reliable} proposed a fault-aware resource allocation approach that applies the price of spot instances, runtime estimation of applications, and task duplication mechanisms to economically run batch jobs in spot instances. Yi et al. \cite{yi2010reducing} proposed checkpointing schemes to reduce the computation price of spot instances and the completion time of tasks. Shastri et al. \cite{shastri2017hotspot} proposed a resource container that enables applications to self-migrate to new spot VMs in a way that optimizes cost-efficiency as the spot prices change. In addition, other researchers worked on modeling spot markets to reduce the spot instance cost and the performance penalty that results from a high number of revocations, by designing optimal bidding strategies \cite{song2012optimal,tang2012towards,zafer2012optimal} and developing prediction schemes \cite{wolski2017probabilistic,wang2018empirical}.

\section{Conclusion}
 
We addressed a challenging problem for applications deployed on cloud spot instances that results from the overhead of employing fault-tolerance mechanisms. We proposed a novel cloud market-based approach that leverages features of cloud spot markets for provisioning spot instances without employing fault-tolerance mechanisms (\texttt{P-SIWOFT}) to reduce the deployment cost and completion time of applications. We evaluated \texttt{P-SIWOFT} in simulations and used Amazon spot instances that contain jobs in Docker containers and realistic price traces from EC2 markets. Our simulation results show that our approach reduces the deployment cost and completion time compared to approaches based on fault-tolerance mechanisms.

\bibliographystyle{IEEEtran}
\bibliography{IEEEabrv,PSIWFT}

\end{document}